\documentclass[aip,sdy,amsmath,amssymb,reprint]{revtex4-2}

\usepackage{dcolumn}
\usepackage{bm}

\usepackage[utf8]{inputenc}
\usepackage[T1]{fontenc}
\usepackage{mathptmx}
\usepackage{etoolbox}

\usepackage{amsmath}
\usepackage[colorlinks = true, linkcolor = blue, citecolor = blue]{hyperref}
\usepackage[mathlines]{lineno}
\usepackage{graphicx}
\usepackage{lipsum}
\usepackage{color,soul}
\usepackage[table]{xcolor}
\usepackage{tabularx} 
\usepackage{tikz}
\usepackage{float}
\usepackage{array}
\usepackage{booktabs}
\usepackage{revsymb}

\makeatletter
\def\@email#1#2{%
\endgroup
\patchcmd{\titleblock@produce}
{\frontmatter@RRAPformat}
{\frontmatter@RRAPformat{\produce@RRAP{*#1\href{mailto:#2}{#2}}}\frontmatter@RRAPformat}
{}{}
}%
\makeatother
\begin{document}

\preprint{AIP/123-QED}

\title[RF acceleration of ultracold electron bunches]{RF acceleration of ultracold electron bunches}

\author{D.F.J. Nijhof*}
\email{d.f.j.nijhof@tue.nl}
\affiliation{Department of Applied Physics and Science Education, Coherence and Quantum Technology Group, Eindhoven University of Technology, P.O. Box 513, 5600 MB Eindhoven, the Netherlands}

\author{T.C.H. de Raadt}
\affiliation{Department of Applied Physics and Science Education, Coherence and Quantum Technology Group, Eindhoven University of Technology, P.O. Box 513, 5600 MB Eindhoven, the Netherlands}

\author{J.V. Huijts}
\affiliation{Department of Applied Physics and Science Education, Coherence and Quantum Technology Group, Eindhoven University of Technology, P.O. Box 513, 5600 MB Eindhoven, the Netherlands}
\affiliation{Institute for Complex Molecular Systems, Eindhoven University of Technology, P.O. Box 513, 5600 MB Eindhoven, The Netherlands}

\author{J.G.H. Franssen}
\affiliation{Doctor X Works BV, 5616 JC Eindhoven, The Netherlands}

\author{O.J. Luiten}
\affiliation{Department of Applied Physics and Science Education, Coherence and Quantum Technology Group, Eindhoven University of Technology, P.O. Box 513, 5600 MB Eindhoven, the Netherlands}
\affiliation{Institute for Complex Molecular Systems, Eindhoven University of Technology, P.O. Box 513, 5600 MB Eindhoven, The Netherlands}
\affiliation{Doctor X Works BV, 5616 JC Eindhoven, The Netherlands}

\date{\today}

\begin{abstract}
	The ultrafast and ultracold electron source, based on laser cooling and trapping of an atomic gas and its subsequent near-threshold photoionization, is capable of generating electron bunches with a high transverse brightness at energies of roughly 10 keV. This paper investigates the possibility of increasing the range of applications of this source by accelerating the bunch using radio-frequency electromagnetic fields. Bunch energies of $\sim35$ keV are measured by analyzing the diffraction patterns generated from a mono-crystalline gold sample. Further analysis points to a largely preserved normalized transverse emittance during acceleration.
\end{abstract}

\maketitle

\section{INTRODUCTION}\label{sec:introduction}

\indent In recent years the ultracold electron source (UCES) has been developed at the Eindhoven University of Technology to produce stable electron bunches at energies up to 10 keV with normalized transverse emittances in the order of a few nm$\cdot$rad at repetition rates of 1 kHz \cite{Claessens2005, Taban_2010, McCulloch2011, Engelen_2013, McCulloch2013, Engelen_2014, Speirs2015a, Franssen_2019_2}.
\\
\indent The bunches extracted from this source are created by the two-step near-threshold photoionization of a laser-cooled and trapped rubidium gas. Their kinetic energy is determined by the static electric field strength in which the laser cooling and trapping occurs. This configuration results in a practical limit for the extraction field potential ($\sim1$ MV/m), caused by the broadening of the laser-cooling transition by breaking the $m_{F}$ degeneracy (Stark shift) \citep{Franssen_2019_2, Gunton_2016}.
\\
\indent Additional acceleration of these bunches is desirable not only because of the increased scope of applications like the investigation of thicker samples and/or sample environments, but also because of the reduced influence of the bunch's self-fields that are partially responsible for bunch quality degradation during free-space propagation \citep{Reiser_2008}.
One way of achieving this is through the use of a resonant radio-frequency (RF) cavity, utilizing a transverse magnetic (TM) mode \citep{Jackson_1998}. In such a mode the electric field is parallel to the propagation direction of the electron bunches, allowing additional acceleration.
\\
\indent Such an RF cavity has been designed at the Coherence and Quantum Technology (CQT) group for the purpose of (longitudinally) compressing high charge electron bunches to sub-ps pulse durations \citep{VanOudheusden2010a}. This cavity, operating at a frequency of 2.99855 GHz (S-band) in the TM$_{010}$ mode can also be used to accelerate ultracold electron bunches. Work presented in this paper will first investigate the viability of this method through \textsc{general particle tracer} \cite{GPT} (GPT, charge particle tracking simulations). Data will then be presented showing that ultracold bunches have been accelerated in this fashion and their final kinetic energy determined through diffraction measurements on a mono-crystalline gold sample. Furthermore, analysis of the diffraction pattern will show that the electron bunch transverse beam quality is (largely) preserved during acceleration.
\\
\indent The structure of this paper is organized as follows: first a brief overview of the UCES is presented in Sec. (\ref{sec:UCES}). In Sec. (\ref{sec:experimental_design}) the electron beamline and RF infrastructure are introduced. Sec. (\ref{sec:particle_tracking_simulations}) will examine the viability of additional acceleration through an RF cavity by investigating the relevant bunch parameters by charged particle simulations. Measurements are then discussed in Sec. (\ref{sec:measurements}) where diffraction patterns generated from a mono-crystalline gold sample are analyzed to calculate the acquired kinetic energy of the bunches. Finally, Sec. (\ref{sec:bunch_quality}) presents additional analysis of the measured diffraction spots, yielding the (transverse) quality of the accelerated electron bunches.
\\
\indent Measurements show that bunches with an initial kinetic energy of $7.3\pm0.1$ keV are accelerated using an RF cavity to energies of $\sim$35 keV with a transverse coherence length of $\sim$1 nm, easily resolving $>$40 diffraction peaks of a mono-crystalline Au sample. Transverse bunch quality estimations based on simulations and experimental data further indicate a transverse normalized emittance of $\sim$6 nm$\cdot$rad.

\section{THE ULTRACOLD ELECTRON SOURCE}\label{sec:UCES}
\begin{figure}[t!]
	\centering		
	\includegraphics[width=1\linewidth]{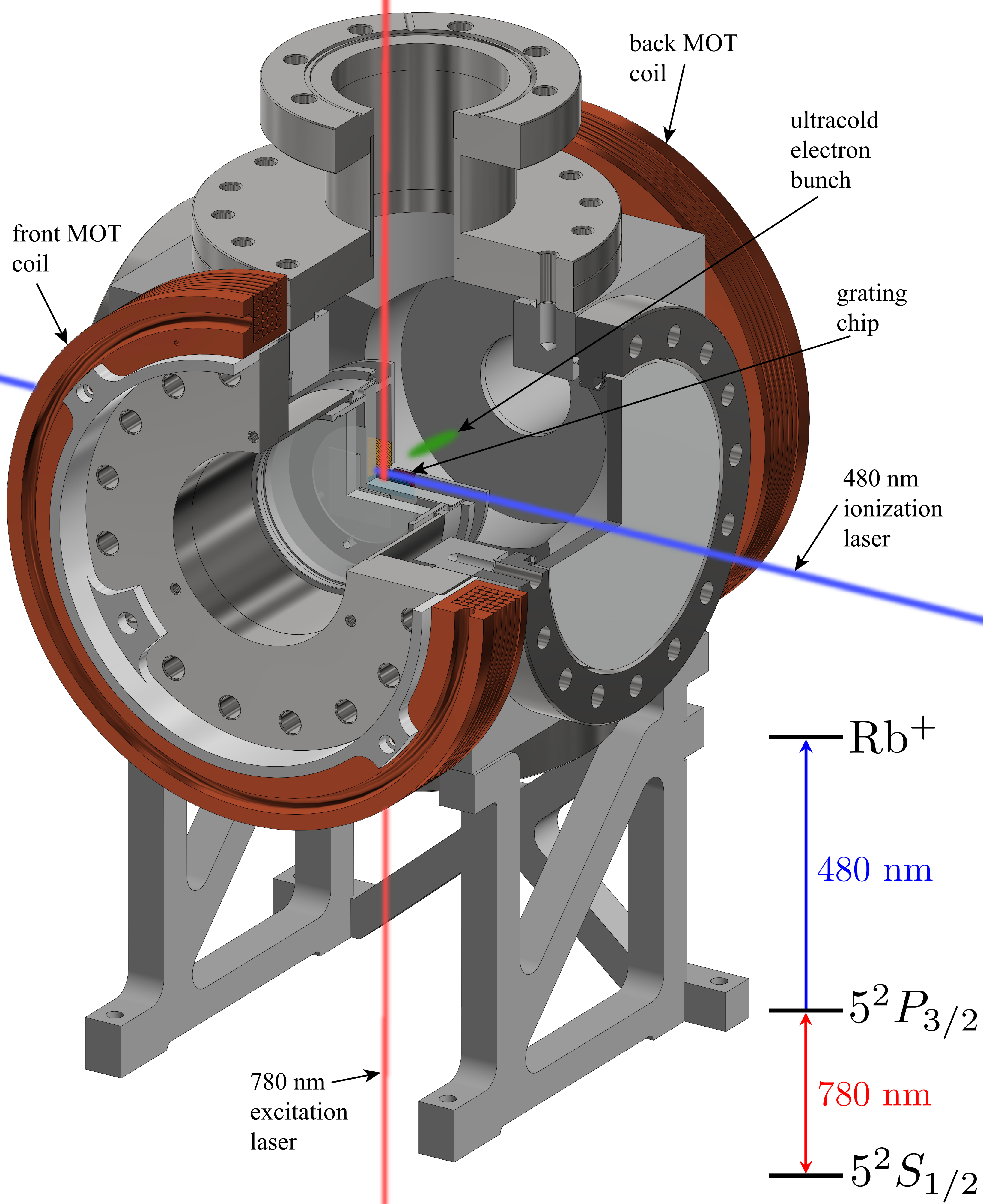}	
	\caption{Schematic representation of the UCES showing the excitation and ionization laser creating an electron bunch (green) which is accelerated downstream.}
	\label{fig:UCES_3_4_beam_scheme}
\end{figure}

\indent The electron source used to create the ultracold electron bunches described in this article works through the laser-cooling and trapping and subsequent (two-step) photo-ionization of an $^{85}$Rb atomic gas. The predecessor of this source consisted of a glass cell and six red-detuned mutually-perpendicular laser beams \citep{Claessens_2007, Taban_2008}. High-quality diffraction patterns were obtained with this source and electron temperatures of $\sim$30 K were found for femtosecond photoionization bunch generation \citep{Engelen_2013, Engelen_2014, Mourik_2014}.
\\
\indent The current ultracold source, shown schematically in Fig. \ref{fig:UCES_3_4_beam_scheme}, retains the same operational principle: rubidium gas is laser-cooled and trapped and electron bunches are extracted using a static field through near-threshold femtosecond photoionization. This section will give a brief description of the source as is relevant for this work.
\begin{figure*}[]
	\centering	
	\includegraphics[width=1\linewidth]{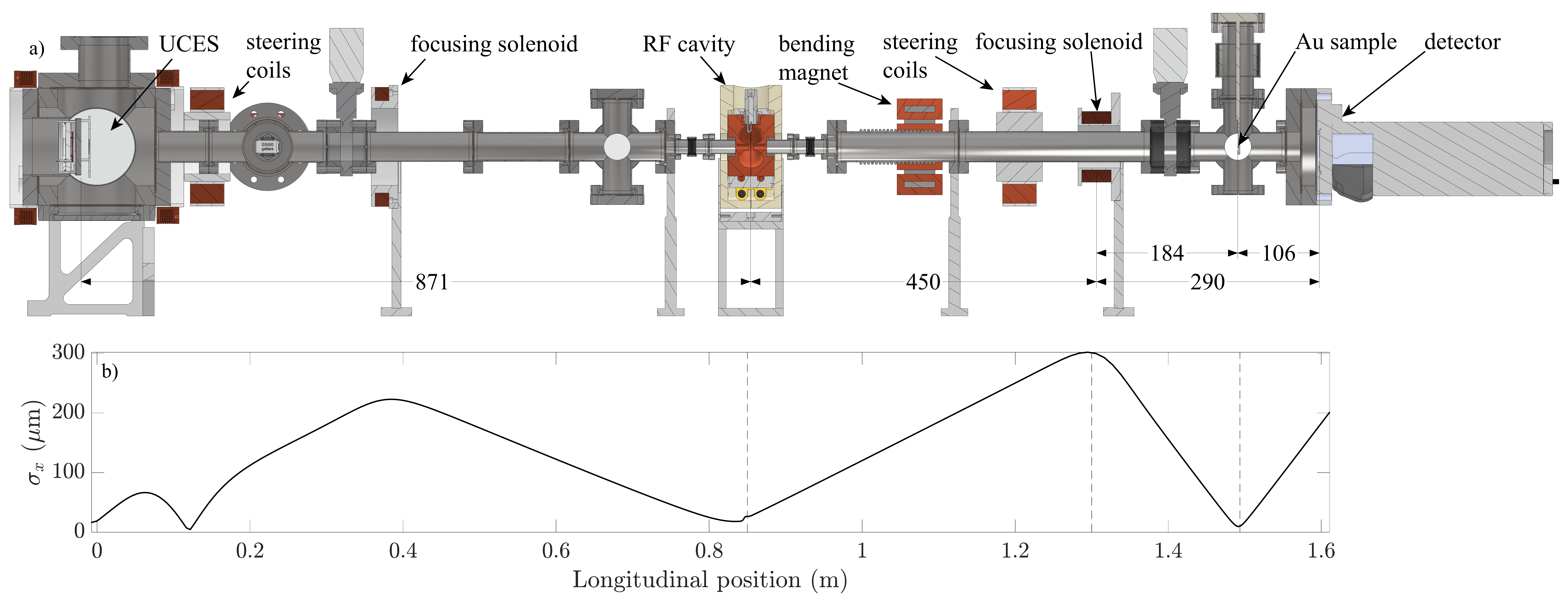}	
	\caption{Schematic representation of the UCES beamline showing the main components (from left to right): the UCES, RF cavity, focusing solenoid, Au sample holder, and the Cheetah T3 detector (units in mm) (a). The rms transverse bunch size as a function of the longitudinal position. The RF cavity, focusing solenoid, and Au sample locations are indicated by dashed vertical lines (b).}
	\label{fig:beamline_schematic}
\end{figure*}
\\
\indent Instead of using three pairs of mutually perpendicular laser beams to realize laser-cooling and trapping, a single circularly polarized red-detuned laser beam (frequency stabilized to the cooling transition $5^{2}S_{1/2}F=3\rightarrow5^{2}P_{3/2}F=4$ of $^{85}$Rb) is used in conjunction with a grating chip to achieve this \citep{Nshii_2013} (for a more detailed description the reader is referred to work by Franssen \& van Ninhuijs \citep{Franssen_2019_2, Ninhuijs_2021}).
\\
\indent To create electron bunches, the trapping laser is turned off for a few \textmu s after which the rubidium atoms are again excited using a 780 nm laser to the $5^{2}P_{3/2}F=4$ state and subsequently ionized by a (wavelength) tunable 480 nm fs laser pulse, propagating perpendicular to the excitation laser, see Fig. \ref{fig:UCES_3_4_beam_scheme}. The wavelength tunability of the ionization lasers enables the generation of electrons with very low excess energies. The laser cooling and trapping, excitation, and ionization is done in a static extraction field of $\sim$1 MV/m at the region of ionization, which results in $\sim$10 keV electron bunches when the transparent cathode is kept at -20 kV.
\\
\indent Typically, electron bunches are created with an initial rms source size of $\sim$30 \textmu m in the transverse and longitudinal direction. At 10 K temperatures this results in rms normalized emittances of $\sim$2 nm$\cdot$rad \citep{Franssen_2019_2}. Recently, pondermotive measurements have been performed in the self-compression point, resulting in measured rms bunch lengths of $735\pm7$ fs \citep{Raadt_2023}.

\section{EXPERIMENTAL DESIGN}\label{sec:experimental_design}

\subsection*{Electron beamline}\label{subsec:beamline}

\begin{figure*}[]
	\centering
	\includegraphics[width=1\linewidth]{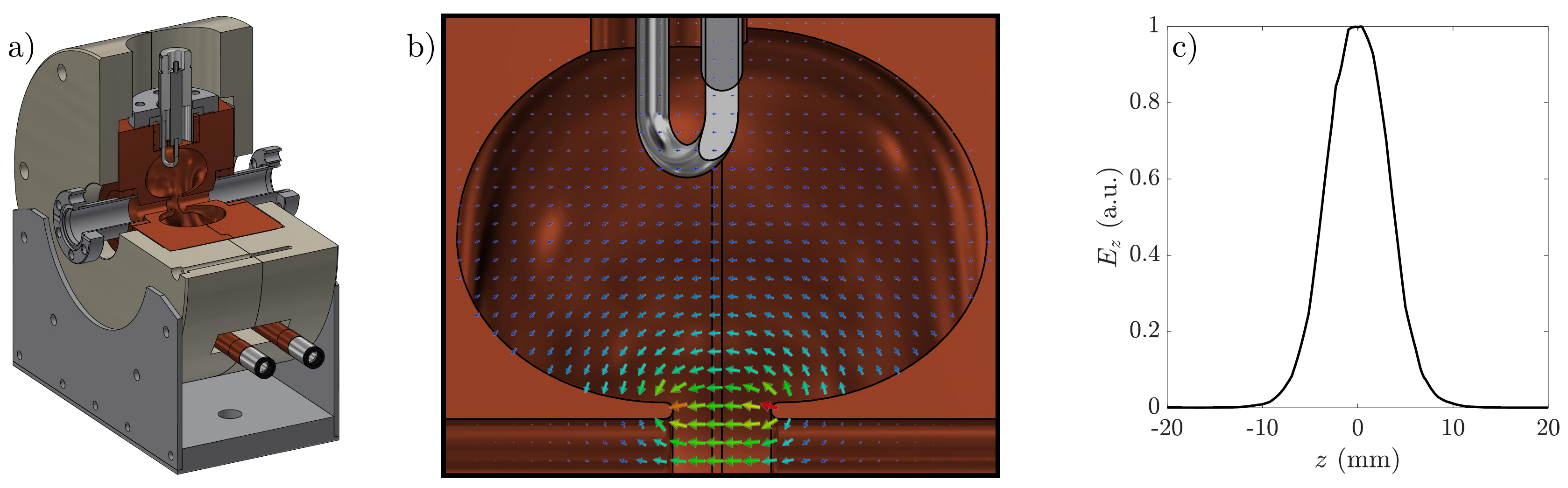}
	\caption{A schematic representation of the TM$_{010}$ cavity (a), a close-up of the cavity's interior with the arrows indicating the electric field vectors in the $x-z$ plane (colors ranging from blue (weak) to red (strong)) (b), the on-axis longitudinal electric field  profile (c).}
	\label{fig:TM010_cavity_schematic_fields}
\end{figure*}

\indent A schematic representation of the beamline used during the experiment is given in Fig. \ref{fig:beamline_schematic}(a), where the UCES is shown, including correction and focusing coils, the TM$_{010}$ cavity, the sample holder which holds the Au sample, and the detector. The detector used in this work is an event-driven Cheetah T3 which is capable of capturing diffraction patterns without the need to block the high intensity central peak \cite{timepix3_2014}. Fig. \ref{fig:beamline_schematic}(b) shows the typical evolution of the transverse rms beam size throughout the beamline with the locations of key components indicated by vertically dashed lines.
\\
\indent The cavity used in this work is shown in Fig. \ref{fig:TM010_cavity_schematic_fields}(a) where the copper structure, the thermally insulating material encasing it, the coupling antenna, and the water-cooling connections are visible. Fig. \ref{fig:TM010_cavity_schematic_fields}(b) shows the top-halve of the copper structure in the $x-z$ plane where the arrows indicate the direction and magnitude of the electric field. Enhancement of the on-axis field strength by the nose-cones can clearly be seen. Fig. \ref{fig:TM010_cavity_schematic_fields}(c) shows the on-axis longitudinal ($E_{z}$) field strength based on \textsc{cst microwave studio} \cite{CST} simulations from (b).
\\
\indent The energy of the RF-accelerated electron bunch is measured through diffraction off a mono-crystalline Au sample.

\subsection*{RF infrastructure}\label{subsec:RF_infrastructure}
\indent Additional acceleration of the $\sim$10 keV ultracold electron bunches is realized through the use of an RF cavity. For this to happen energy needs to be stored in the cavity in the form of electromagnetic fields. In this work a pulsed operation of the accelerating cavity is used: RF pulses with a duration of 10 \textmu s were used to fill the structure with the required energy. The generation of the RF signal is realized as follows: a Ti:Saph-based mode-locked oscillator (Coherent MANTIS oscillator) is used as the master clock of the experiment, operating at 74.9625 MHz.
This signal is used to synchronize a 2.99855 GHz electronic oscillator (40$^{\textnormal{th}}$ harmonic) to the laser system \citep{Kiewiet_2002}. 
\\
\indent The phase-stabilized $\sim$3 GHz continuous-wave (CW) signal is used as the input for the generation of the 10 \textmu s RF pulses. 
\\
\indent The cavity is designed to operate at $f_{0}=2.99855$ GHz with a (measured) 
unloaded quality factor of $Q_{0}=8960$ \citep{Oudheusden2010}. More information on the generation of the RF pulse and the characterization of the cavity can be found in the supplementary material.
\section{PARTICLE TRACKING SIMULATIONS}\label{sec:particle_tracking_simulations}

\begin{figure*}[t!]
	\centering
	\includegraphics[width=0.9\linewidth]{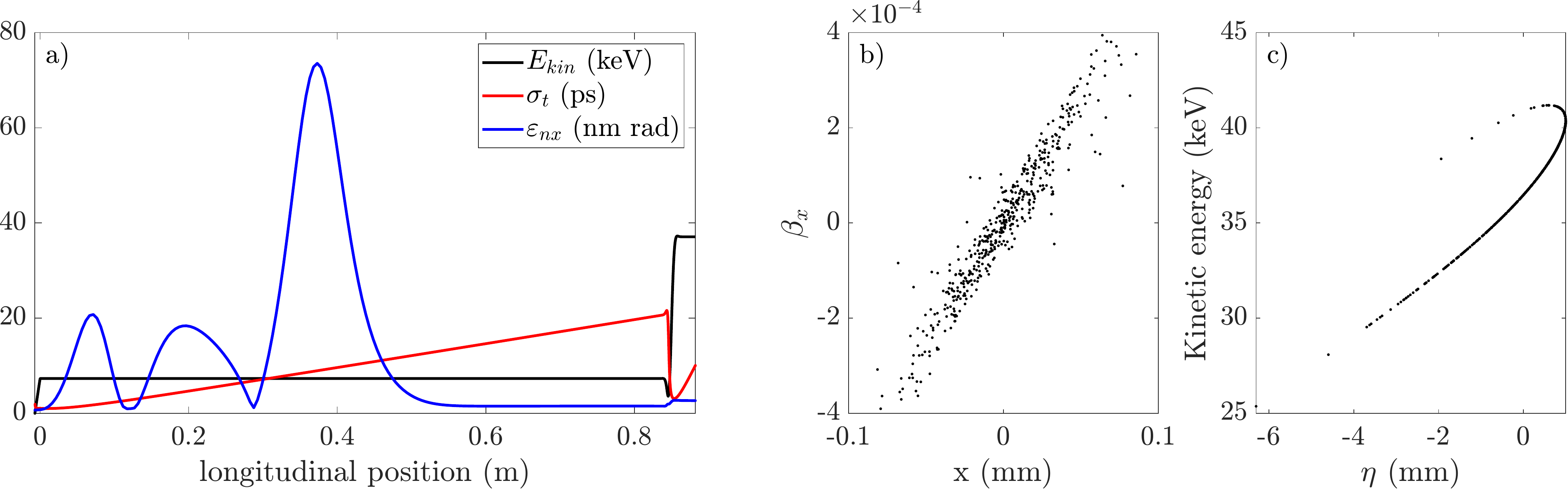}
	\caption{GPT simulations of a 0.1 fC bunch showing: the kinetic energy (black), rms temporal length (red), and the normalized transverse emittance in the $x$-direction (blue) as a function of the longitudinal position (a), the transverse phase-space ($x$-$\beta_{x}$) (b), and the kinetic energy as a function of the relative longitudinal coordinate $\eta$ (c).}
	\label{fig:bunch_acceleration_simulation}
\end{figure*}

\indent \textsc{general particle tracer} \citep{GPT} simulations are done with realistic beamline parameters to investigate the bunch's behavior. Throughout the experiments the applied potential to the static accelerator was set to -16 kV and the laser-cooled and trapped atom cloud was created $\sim$7 mm in front of the grating surface. The excitation and ionization lasers had rms spot sizes of $\sim$15 \textmu m, creating $\sim$0.1 fC bunches. At electron temperatures of 10 K, a thermal emittance of $\sim$0.6 nm$\cdot$rad is found.
\\
\indent These parameters result in bunches leaving the UCES at approximately 7.3 keV, based on realistic field maps, as shown in Fig. \ref{fig:bunch_acceleration_simulation}(a), along with the rms temporal length $\sigma_{t}$ and normalized transverse emittance $\varepsilon_{nx}$ as a function of the longitudinal position $z$.
\\
\indent As the bunches drift towards the accelerated structure they diverge in the longitudinal direction after exiting the source due to the imparted energy spread on the bunch. This energy spread is directly related to the dimensions of the overlap of the excitation and ionization laser in the overlap region; electrons created at the back of the bunch are accelerated to higher energies compared to electrons created at the front. This results in a longitudinal self-compression point just behind the exit of the source, after which the bunch diverges.
\\
\indent The bunch arrives at the compression cavity with an rms pulse length of $\sim$18 ps. The pulse length of the bunch, combined with a low kinetic energy and a 333 ps period of the RF accelerator structure results in non-homogeneous acceleration of the bunch (this is mainly due to the long transit time of the bunches in the cavity, which was originally designed for 100 keV bunches). The transverse/longitudinal phase space of the bunch exiting the cavity is shown in Fig. \ref{fig:bunch_acceleration_simulation}(b/c), where $\eta$ represents the relative longitudinal position of each electron in the bunch with respect to their average longitudinal position. These representations of the bunch are obtained just behind the exit of the RF accelerator structure and show the sub-optimal longitudinal phase space of the bunch. An average kinetic energy in the range of 40 keV is expected when a forward power of 500 W is supplied to the RF structure.
\\
\indent During the drift to the RF accelerator structure the normalized transverse emittance grows to $\sim$1.5 nm$\cdot$rad, and right after acceleration an increase to $\sim2.7$ nm$\cdot$rad is found. 

\section{KINETIC ENERGY MEASUREMENTS}\label{sec:measurements}

\begin{figure}[t!]
	\centering
	\includegraphics[width=1\linewidth]{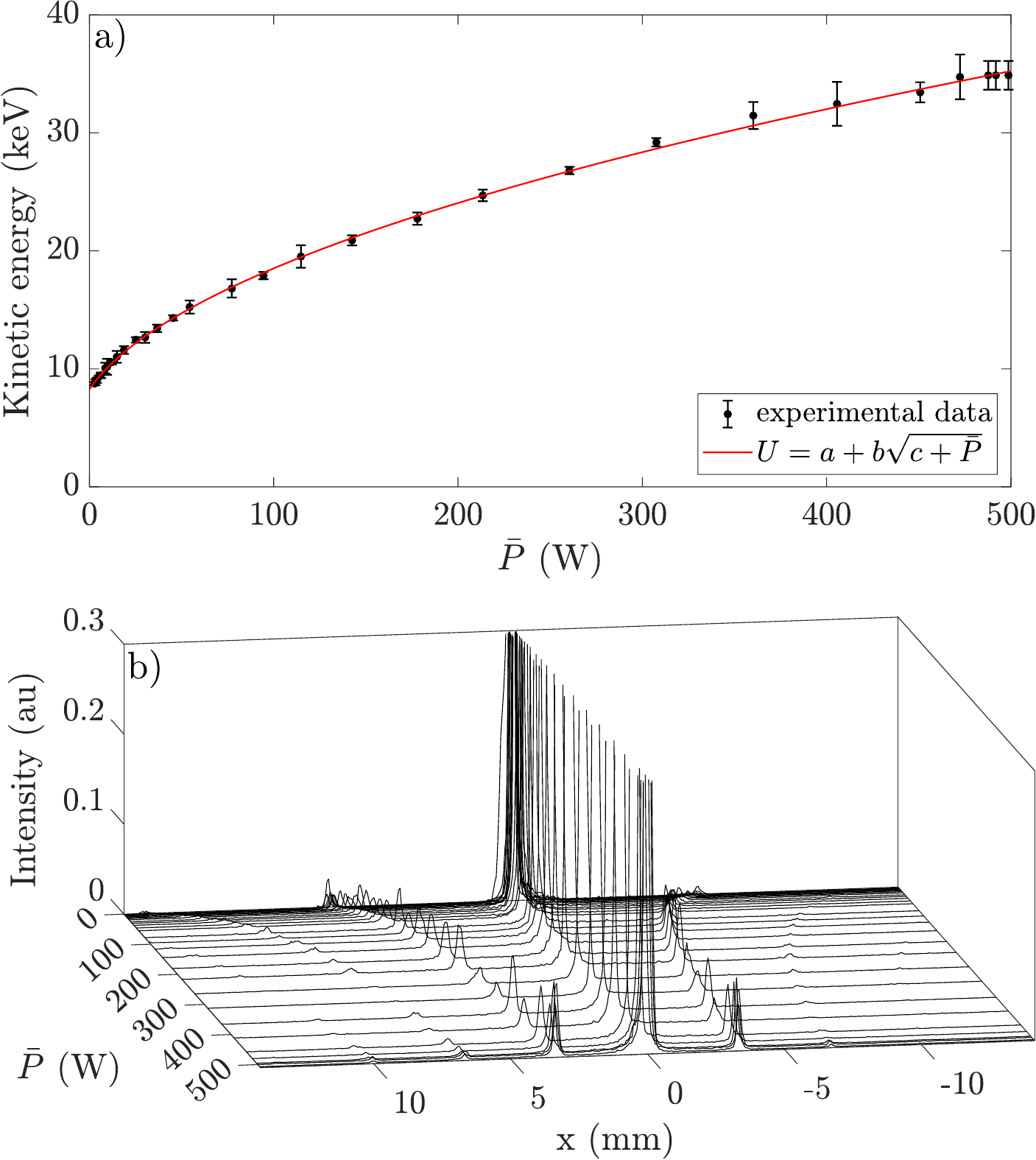}
	\caption{The determined kinetic energy of the accelerated bunches as a function of the RF power supplied to the cavity (black data points) fitted with a basic square-root function (solid red) with $a=3.80$, $b=1.38$, and $c=9.14$ (a). 3D plot showing a line-out profile (along the center) of the detector image as a function of the power supplied to the RF structure (b).}
	\label{fig:accelerated_diffraction_kinetic_energy_summed_profiles}
\end{figure}

\begin{figure}[t!]
	\centering
	\includegraphics[width=1\linewidth]{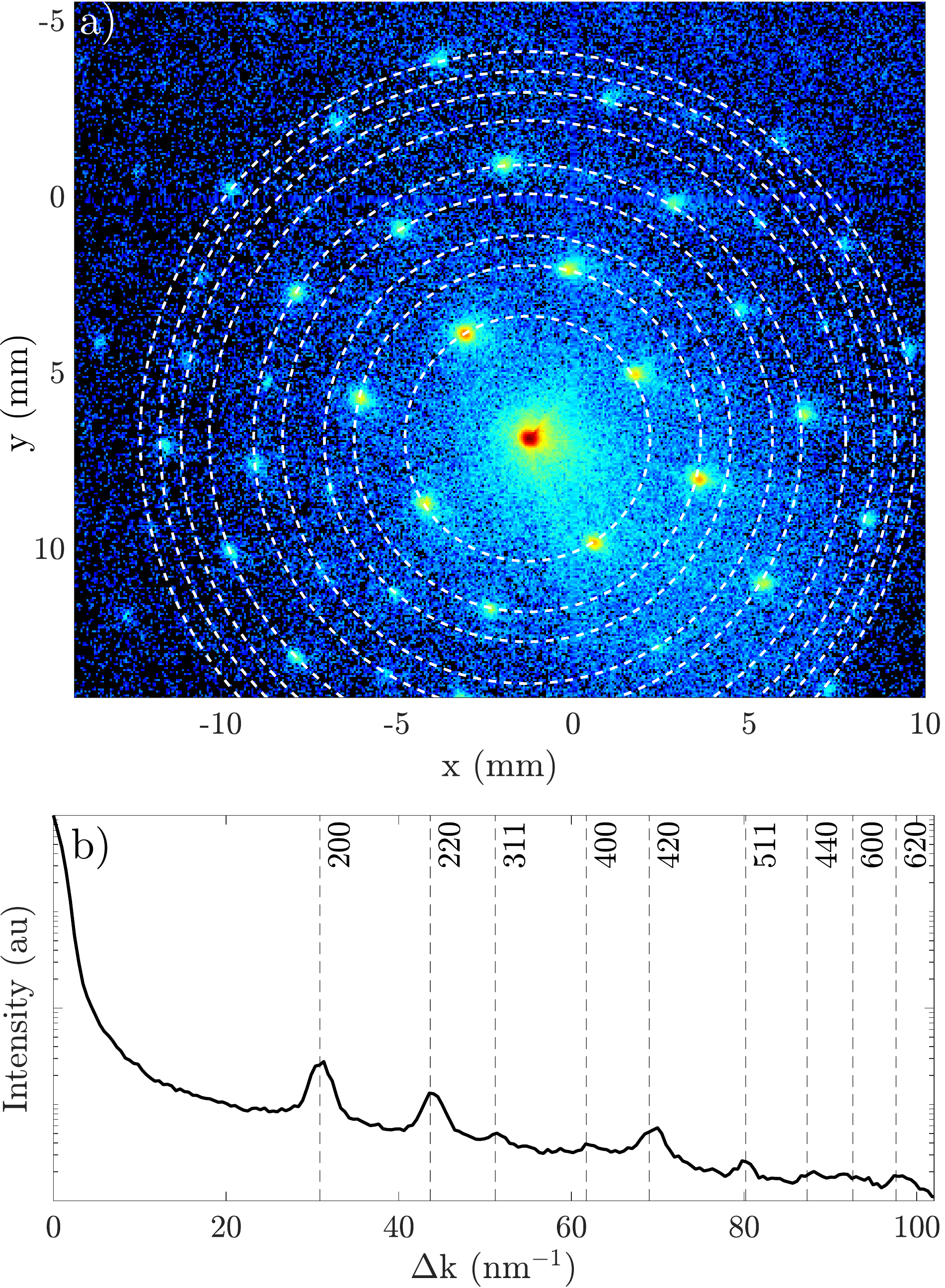}
	\caption{Diffraction pattern collected on the Cheetah T3 detector, generated by 32.6 keV electrons (412 W supplied to accelerator structure) on Au, the dashed rings indicate the groups of Miller indices listed in Table. (\ref{table:Au_parameters}) (a). The corresponding $\Delta k$-space showing the azimutally averaged intensity with the theoretically calculated recpirocal lattice planes $q_{hkl}$ indicated by dashed lines (b).}
	\label{fig:accelerated_diffraction_hkl_and_k_space}
\end{figure}

\indent First, measurements are conducted without delivering power to the RF accelerator structure, i.e. only DC acceleration to $\sim$7 keV. The bunch is focused on a mono-crystalline Au sample positioned at a distance of $\sim$106 mm from the detector. The crystal structure of Au is known to be face-centered cubic with a lattice parameter of 4.07\AA$\;$\citep{Au_springer_2023}, the crystal is grown in a (100) orientation, giving familiar lattice spacings.
\\
\indent An expression for the longitudinal momentum of the bunch as a function of the separation between the 1$^{\textnormal{st}}$ order diffracted peaks and the $0^{\textnormal{th}}$ order peak (henceforth called the direct beam) can be found by substituting the expression for the de Broglie wavelength in Bragg's diffraction law and rewriting in terms of $\gamma\beta$:
\begin{equation}\label{eq:debroglie_and_bragg}
	m_{e}c\gamma\beta=\frac{nh}{2d_{hkl}\sin\theta_{B}},
\end{equation}
\noindent where $d_{hkl}$ is the spacing between parallel lattice planes with Miller indices $hkl$, $\theta_{B}$ the Bragg angle, $n$ the order of diffraction, $h$ Planck's constant, $m_{e}$ the electron rest mass, $c$ the speed of light, $\gamma=\frac{1}{\sqrt{1-\beta^{2}}}$ the Lorentz factor, and $\beta=\frac{v}{c}$ the normalized bunch velocity.
\\
\indent An overview of the relevant crystal parameters is given in Table (\ref{table:Au_parameters}). Here the peak positions are given by $q_{hkl}=\frac{2\pi}{a}\sqrt{h^2+k^2+l^2}$ and the spacing between parallel lattice planes $d_{hkl}=\frac{2\pi}{q_{hkl}}$.
\begin{table}[b!]
	\caption{Relevant theoretical parameters for diffraction phenomena observed during measurements: the Miller indices $hkl$, the peak positions $q_{hkl}$, and the parallel lattice spacings $d_{hkl}$.}
	\begin{tabularx}{\linewidth}{@{\hskip 0.5cm}c @{\hskip 2.1cm} c @{\hskip 1.7cm} c}
		\toprule
		h k l & $q_{hkl}$ (nm$^{-1}$) & $d_{hkl}$ (pm)\\
		\hline\hline			  
		2 0 0 & 30.81 & 203.91 \\
		2 2 0 & 43.58 & 144.19 \\
		3 1 1 & 51.10 & 122.96 \\
		4 0 0 & 61.63 & 101.95 \\
		4 2 0 & 68.90 & 91.19  \\
		5 1 1 & 80.06 & 78.49  \\
		4 4 0 & 87.45 & 72.09  \\
		6 0 0 & 92.44 & 67.97  \\
		6 2 0 & 97.44 & 64.48  \\
		\bottomrule
	\end{tabularx}
	\label{table:Au_parameters}
\end{table}
\begin{figure}[t!]
	\centering
	\includegraphics[width=1\linewidth]{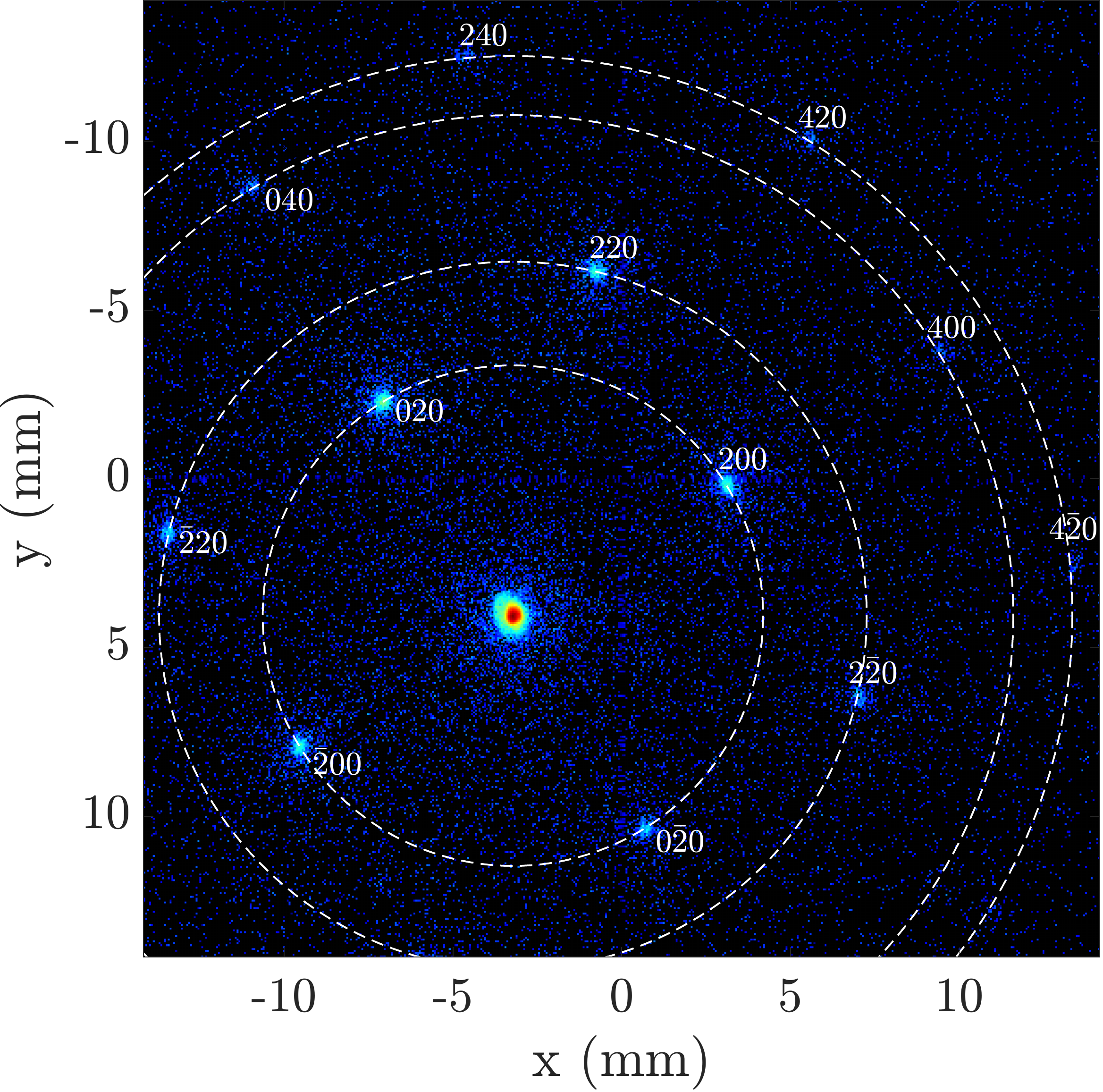}
	\caption{Detector image of the diffraction pattern of a mono-crystalline Au sample produced by the electron bunches extracted from the UCES without additional RF acceleration. The numbers indicate the Miller indices $hkl$ corresponding to the observed diffraction peak with the dashed line indicating the diffraction rings that would be obtained if the sample had been polycrystalline.}
	\label{fig:initial_kinetic_energy}
\end{figure}
\\
\indent With the distance between the gold sample and the detector plane known, the kinetic energy can be calculated from the measured separation between the direct beam and 1$^{\textnormal{st}}$ order diffracted spots. During the measurement series, the seed RF power is subsequently increased in steps of 1 dB. Fig. \ref{fig:initial_kinetic_energy} shows the summed diffraction pattern consisting of 10 separate measurements each containing 100 shots, obtained at a repetition frequency of 1 kHz. The color scale is logarithmic to increase the visibility of the low intensity peaks. Here the first order diffraction peaks are clearly visible, indicated in white by their corresponding $hkl$ values. From this measurement the initial kinetic energy of the ultracold electron bunches is determined to be $7.3\pm0.1$ keV through Eq. (\ref{eq:debroglie_and_bragg}), corresponding to the expected initial bunch energy based on field maps and operational parameters.
\\
\indent Results of the measurements are presented in Fig. \ref{fig:accelerated_diffraction_kinetic_energy_summed_profiles}(a), which shows the determined kinetic energy of the electron bunches as a function of the power delivered to the cavity. The data points represent the average of the determined kinetic energies of the combined $020$ and $0\bar{2}0$ diffraction spots, the error bars give the standard deviation in the obtained energy from those spots. In red a square-root fit shows excellent agreement with the measured data.
\\
\indent Fig. \ref{fig:accelerated_diffraction_kinetic_energy_summed_profiles}(b) shows the normalized line-out profile (along the horizontal axis) of detector images and displayed in a 3D plot as a function of the power supplied to the structure. Here it can clearly be seen that diffraction spots move toward the direct beam  as the power delivered to the RF cavity increases. 
\\
\indent Fig. \ref{fig:accelerated_diffraction_hkl_and_k_space}(a) shows a diffraction pattern obtained with $\sim32.6$ keV electron bunches (100 shots, 0.1 s exposure). Many more diffraction peaks have become visible on the detector compared to the image shown in Fig. \ref{fig:initial_kinetic_energy}. This is a result of the decreasing wavelength of the electrons and a corresponding increase of the radius of the Ewald sphere. 
Fig. \ref{fig:accelerated_diffraction_hkl_and_k_space}(b) shows the azimutally averaged intensity plot of the corresponding figure above it. The dashed lines correspond to the diffraction peak groups that were obtained from theory ($q_{hkl}$ in Table \ref{table:Au_parameters}) and $\Delta k$ is obtained through $\Delta k=2\left|\mathbf{k}_{0}\right|\sin\theta_{B}$ with data from the detector image. 

\section{BUNCH QUALITY}\label{sec:bunch_quality}

\begin{figure}[t!]
	\centering
	\includegraphics[width=1\linewidth]{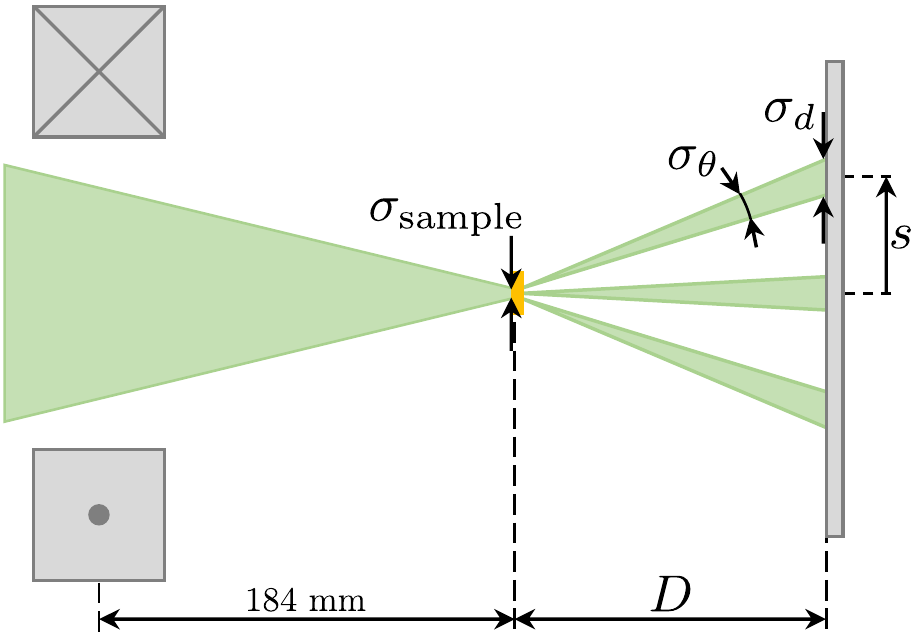}
	\caption{A schematic representation of the diffraction experiment showing relevant parameters used in the determination of the transverse coherence length $L_{T}$.} 
	\label{fig:schematic_diffraction}
\end{figure}

\begin{figure*}[t!]
	\centering
	\includegraphics[width=0.9\linewidth]{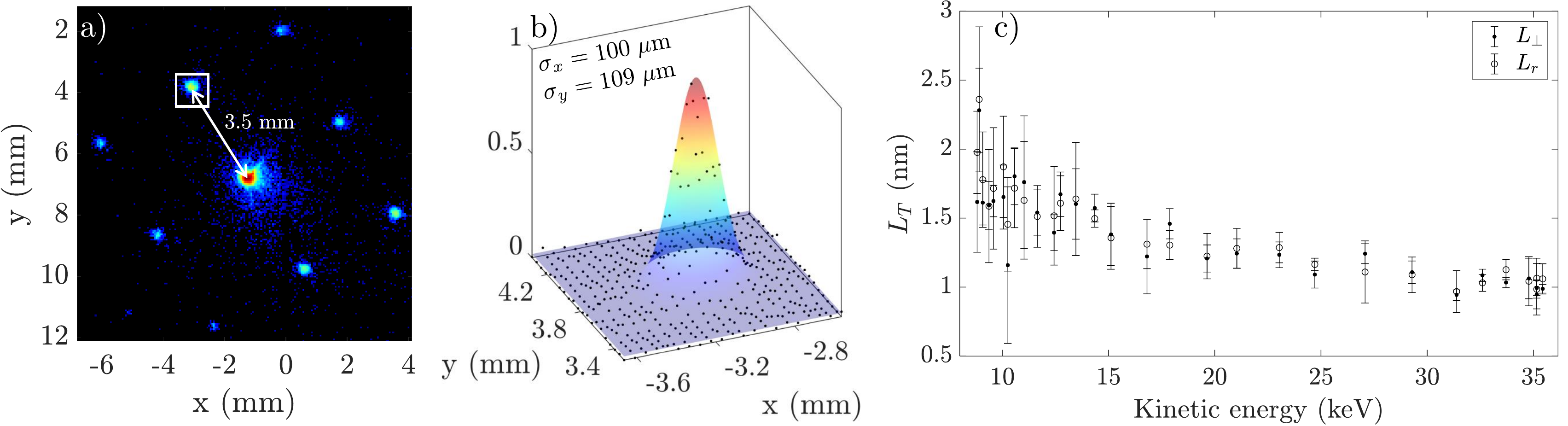}
	\caption{Routine for determining the transverse coherence length showing, a portion of the diffraction pattern presented in Fig. \ref{fig:accelerated_diffraction_hkl_and_k_space} showing the separation between the unscattered electrons and the first order diffraction peak (a), a 2-dimensional Gaussian surface fit through the (200) diffraction spot highlighted by the white box (b), and the resulting calculated transverse coherence lengths in the $x$ and $y$-direction as calculated by Eqn. (\ref{eq:Ltrans_final}), as a function of the calculated bunch energies according to Eqn. (\ref{eq:debroglie_and_bragg}) (c).}
	\label{fig:transversecoherencelengthfigure}
\end{figure*}

\indent From the diffraction patterns obtained in Sec. (\ref{sec:measurements}), the transverse coherence length $L_{T}$ of the electron bunch at the sample plane can be calculated via:
\begin{equation}\label{eq:Ltrans}
	L_{T}\equiv\frac{\hbar}{\sigma_{p_{x}}}=\frac{\lambda_{B}}{2\pi\sigma_{\theta}},
\end{equation}
where $\lambda_{B}=\frac{h}{m_{e}c\gamma\beta}$ is the de Broglie wavelength and $\sigma_{\theta}$ the rms angular divergence of the electron bunch at the sample plane. During the diffraction experiments the transverse beam waist is repositioned on the Au sample (by varying the current through the focusing solenoid) each time the power supplied to the RF cavity is increased. By doing this the diffraction spots will diverge onto the detector enlarged, but are still distinguishable \citep{Mourik_2014}, as is seen in Fig. \ref{fig:accelerated_diffraction_hkl_and_k_space}(a). This allows analysis of the detector image under the assumption that the rms angular divergence ($\sigma_{\theta}$) dominates the diffracted spot sizes. For this work this criterion becomes $\sigma_{\theta}\gg\frac{\sigma_{\textnormal{sample}}}{D}\approx0.18$ where $\sigma_{\textnormal{sample}}$ is the rms spot size on the Au sample and $D$ the separation between the Au sample and the detector plane. 
\\
\indent Eqn. (\ref{eq:Ltrans}) can be re-written by substituting the diffraction angle obtained from Bragg's law: $\gamma\beta\approx \frac{h}{2d_{hkl}\theta_{B} m_{e}c}$ under the paraxial approximation, here $d_{hkl}\approx2.04$\AA$\;$is the shortest distance between two crystal planes of Au for which diffraction occurs (see Table. (\ref{table:Au_parameters})). The transverse coherence length can then be expressed as $L_{T}\approx \frac{d_{hkl}\theta_{B}}{\pi\sigma_{\theta}}$. Finally, the substitutions $\theta_{B}\approx \frac{s}{2D}$ with $s$ the separation between the direct beam and the (200) diffraction spots, and $\sigma_{\theta}\approx\frac{\sigma_{d}}{D}$ with $\sigma_{d}$ the rms spot-size of the diffracted spots on the detector are made, resulting in the following expression for the transverse coherence length:
\begin{equation}\label{eq:Ltrans_final}
	L_{T}\approx\frac{d_{hkl}s}{2\pi\sigma_{d}}.
\end{equation}
With this expression $L_{T}$ can be determined by analyzing portions of diffraction patterns like the one shown in Fig. \ref{fig:accelerated_diffraction_hkl_and_k_space}(a). In this configuration, the analysis can be extended to estimate the normalized transverse emittance of the bunch at the sample plane:
\begin{equation}\label{eq:emittance}
	\varepsilon_{n}=\frac{\lambdabar}{C_{T}},
\end{equation}
where $\lambdabar=\frac{\hbar}{m_{e}c}\approx0.39$ pm is the reduced Compton wavelength and $C_{T}=\frac{L_{T}}{\sigma_{\textnormal{sample}}}$ the relative transverse coherence at the sample.
\\
\indent A schematic representation of the diffraction experiment is shown in Fig. \ref{fig:schematic_diffraction}. The transverse coherence length is now calculated for every composite diffraction pattern (consisting of 100 shots taken during 0.1 s exposures), of which three are collected per RF power setting. First the separation $s$ between the diffraction spots and the direct beam is determined as is shown in Fig. \ref{fig:transversecoherencelengthfigure}(a). These spots are subsequently fitted with a 2D Gaussian intensity profile, from which the rms widths $\sigma_{\perp}$ and $\sigma_{r}$ are determined (Fig. \ref{fig:transversecoherencelengthfigure}(b)). This is done for the $020$, $200$, and $0\bar{2}0$ diffraction spots (see Fig. \ref{fig:initial_kinetic_energy}).
\\
\indent As the energy spread affects the bunch profiles in the radial direction $r$, the bunch dimensions whose directions are perpendicular ($\perp$) to the radial coordinate are treated separately. From this data it is found that the average rms spot sizes for all RF attenuation settings are $\sigma_{\perp}=118\pm13\;\mu$m and $\sigma_{r}=118\pm18\;\mu$m respectively. This results in a corresponding rms angular spread on the sample of $\sim$1.1 mrad in both the radial and transverse direction, satisfying the requirement that $\sigma_{\theta}\gg0.18$ mrad.
\\
\indent From this data the transverse coherence lengths are calculated, shown in Fig. \ref{fig:transversecoherencelengthfigure}(c). The error bars give the standard deviation in these data points per RF attenuation setting. It is found that as the bunches are accelerated to higher energies, the transverse coherence length converges to $L_{T}\approx1$ nm.
\\
\indent The somewhat larger values for $L_{T}$ at smaller bunch energies can be attributed to the positioning of the transverse waist. In the supplementary material it will be shown that, during the experiment, lower energetic bunches have a transverse waist before the Au sample plane. This results in a larger drift distance to the detector, and subsequently a larger value in the numerator of Eqn. (\ref{eq:Ltrans_final}).
\begin{figure}[b!]
	\centering
	\includegraphics[width=1\linewidth]{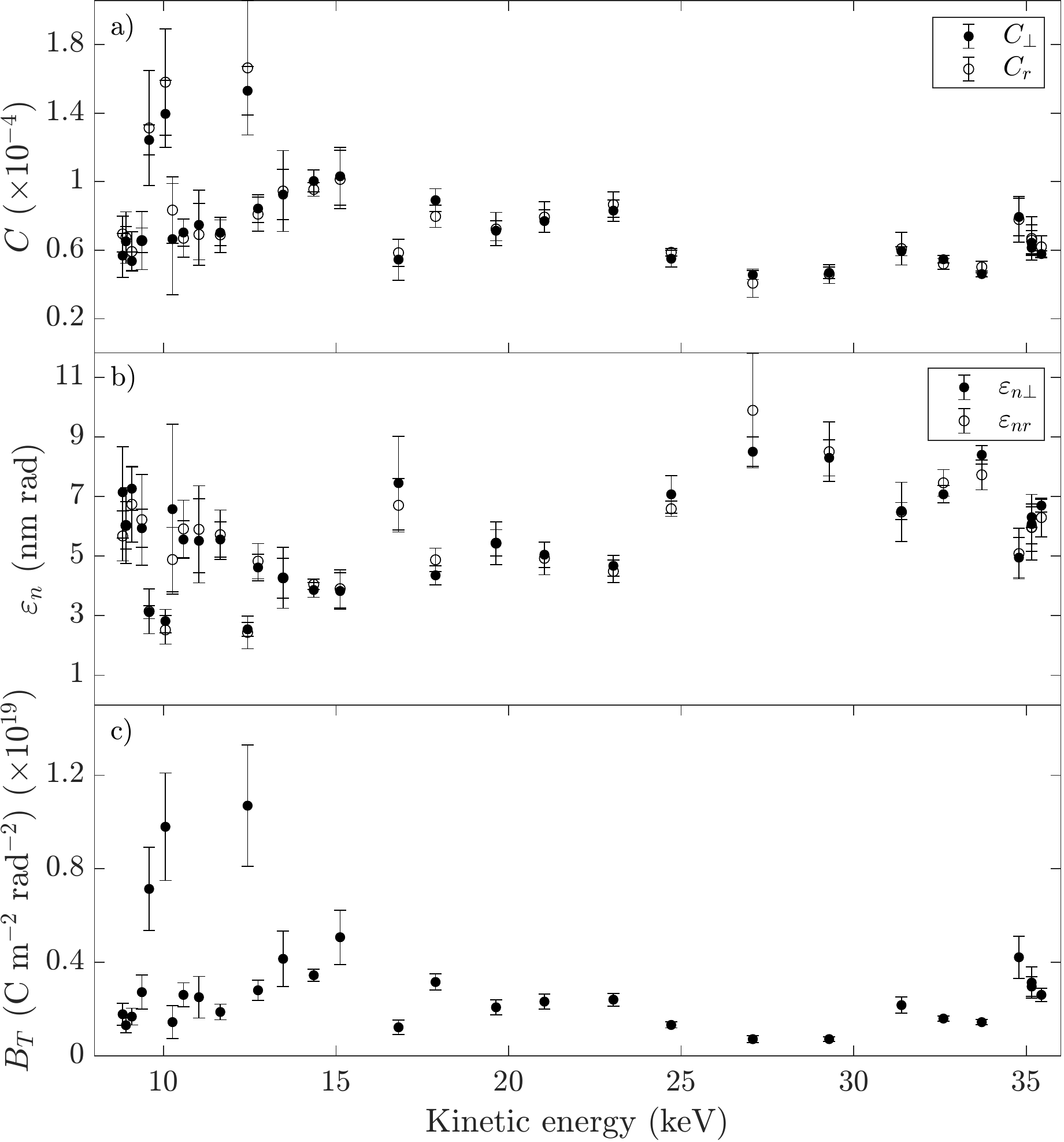}
	\caption{Bunch parameters based on a combination of the measured transverse coherence length and simulated bunch properties, showing: the relative transverse coherence (a), the normalized transverse emittance (b), and the reduced transverse brightness (c).}
	\label{fig:coherenceemittancebrightness}
\end{figure}
\\
\indent 
The relative coherence at the location of the Au sample is shown in Fig. \ref{fig:coherenceemittancebrightness}(a) as a function of the measured kinetic energy of the bunches. The rms spot size on the Au sample is obtained through GPT simulations and was found to be $19\pm6\;\mu$m when averaged over all powers supplied to the RF structure.
\\
\indent Fig. \ref{fig:coherenceemittancebrightness}(b) shows the calculated normalized transverse emittance as calculated by Eqn. (\ref{eq:emittance}). The initial normalized `thermal' emittance of the bunch at the source is determined by the electron temperature and the volume from which they are extracted. The rms transverse spot sizes of the ionization and excitation laser were measured to be $\sigma_{0}\approx15\;\mu$m. These lasers cross perpendicularly, which defines the ionization `overlap' volume and thus the source size. With a perfect overlap between the two lasers a thermal emittance of $\varepsilon_{n,\textnormal{thermal}}=\sigma_{0}\sqrt{\frac{k_{b}T}{m_{e}c^{2}}}\approx0.6$ nm$\cdot$rad is found for 10 K electron temperatures. The bunch quality degrades as the bunch's self fields and non-linear forces from focusing optics affect the phase-space distribution. GPT simulations indicate an emittance growth to $\sim$1.5 nm$\cdot$rad just before the bunch enters the RF accelerator. With an optimized transverse bunch size and an optimal initial phase of the RF structure the normalized emittance grows to approximately 2.7 nm$\cdot$rad during RF acceleration. The measured normalized transverse emittance is larger than in simulations, suggesting a sub-optimal injection in the structure.
\\
\indent Finally, the reduced transverse brightness $B_{T}=\frac{Q}{\varepsilon_{nr}\varepsilon_{n\perp}}$ is plotted in Fig. \ref{fig:coherenceemittancebrightness}(c), obtained by combining Fig. \ref{fig:coherenceemittancebrightness}(b) and \ref{fig:GPT_quality_checks}(c). The average amount of charge per RF setting reaching the detector is measured by the detector itself (see supplementary material). With the normalized transverse emittances determined via Eqn. (\ref{eq:emittance}) the reduced transverse brightness is found to be rather constant, especially considering the relatively large error bars found for higher values of $B_{T}$.

\section{CONCLUSIONS AND OUTLOOK}\label{sec:conclusion} 
\indent Ultracold electron bunches produced through photo-ionization of a laser-cooled atomic gas and extracted by a static field to $7.3\pm0.1$ keV have been accelerated to higher energies through the use of a resonant 2.99855 GHz RF cavity operating in the TM$_{010}$ mode. Measurements of electron diffraction on a single-crystal gold target have shown electron-bunch energies of $\sim$35 keV, corresponding well with expectations obtained from detailed particle tracking simulations. 
\\
\indent Acceleration to even higher kinetic energies can be easily realized with commercially available high power solid-state amplifiers. Incorporation of a bunching-cavity prior to accelerating the bunches should result in higher longitudinal and transverse coherence of the electron bunches and is desirable for further experiments like protein crystallography \citep{Nijhof_2023}$^{\textnormal{ (forthcoming)}}$.
\\
\indent The quality of the individual diffraction patterns implies that transverse bunch quality is largely preserved as the kinetic energy is tuned by varying the RF power to the accelerator structure, implying an energy-tunable ultra-cold electron source. Transverse coherence lengths are found to be $\sim$1 nm when the bunch is accelerated to $\sim$35 keV. Combining the measured values for $L_{T}$ with particle tracking simulations resulted, under various assumptions, in normalized emittances of $\sim$6 nm$\cdot$rad. This value is relatively large compared to earlier measurements \citep{Franssen_2019_2} and is attributed to sub-optimal injection of the electron bunches in the RF structure. Furthermore, the measurement series was not initially intended to yield the bunch quality and the setup was thus not optimized for this.

\section*{Acknowledgments}

The author would like to thank E. Rietman, H. van Doorn, and H. van den Heuvel for their technical support and expertise. This publication is part of the project ColdLight: From laser-cooled atoms to coherent soft X-rays (with project number 741.018.303 of the research programme ColdLight) which is (partly) financed by the Dutch Research Council (NWO). J.V. Huijts is supported by a MSCA Fellowship of the European Union as well a Branco Weiss Fellowship - Society in Science, administered by the ETH Zürich.

\section*{Data availability}

The data that support the findings of this study are available from the corresponding author upon reasonable request.

\section*{References}

\bibliography{UCES_acceleration_library.bib}

\section{Supplementary material}\label{sec:supplementary_material}

\subsection{RF signal generation}

\begin{figure}[b!]
	\centering
	\includegraphics[width=1\linewidth]{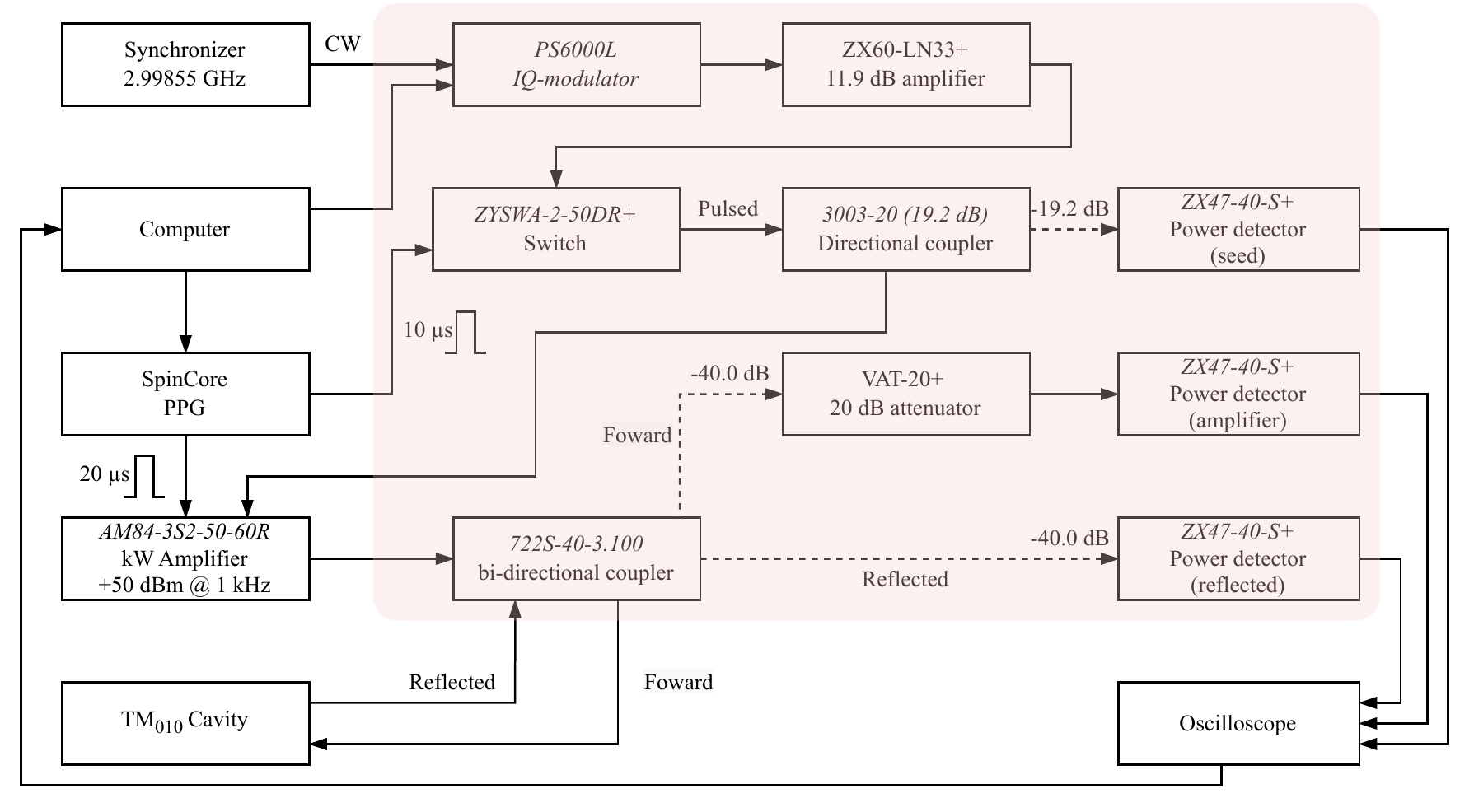}
	\caption{A schematic representation of the components used to generate 10 \textmu s RF pulses to feed into the TM$_{010}$ acceleration cavity, the dashed lines are indicative of the low-power signals split off from the (bi-) directional couplers.}
	\label{fig:DVD}
\end{figure}
\begin{figure}[t!]
	\centering
	\includegraphics[width=1\linewidth]{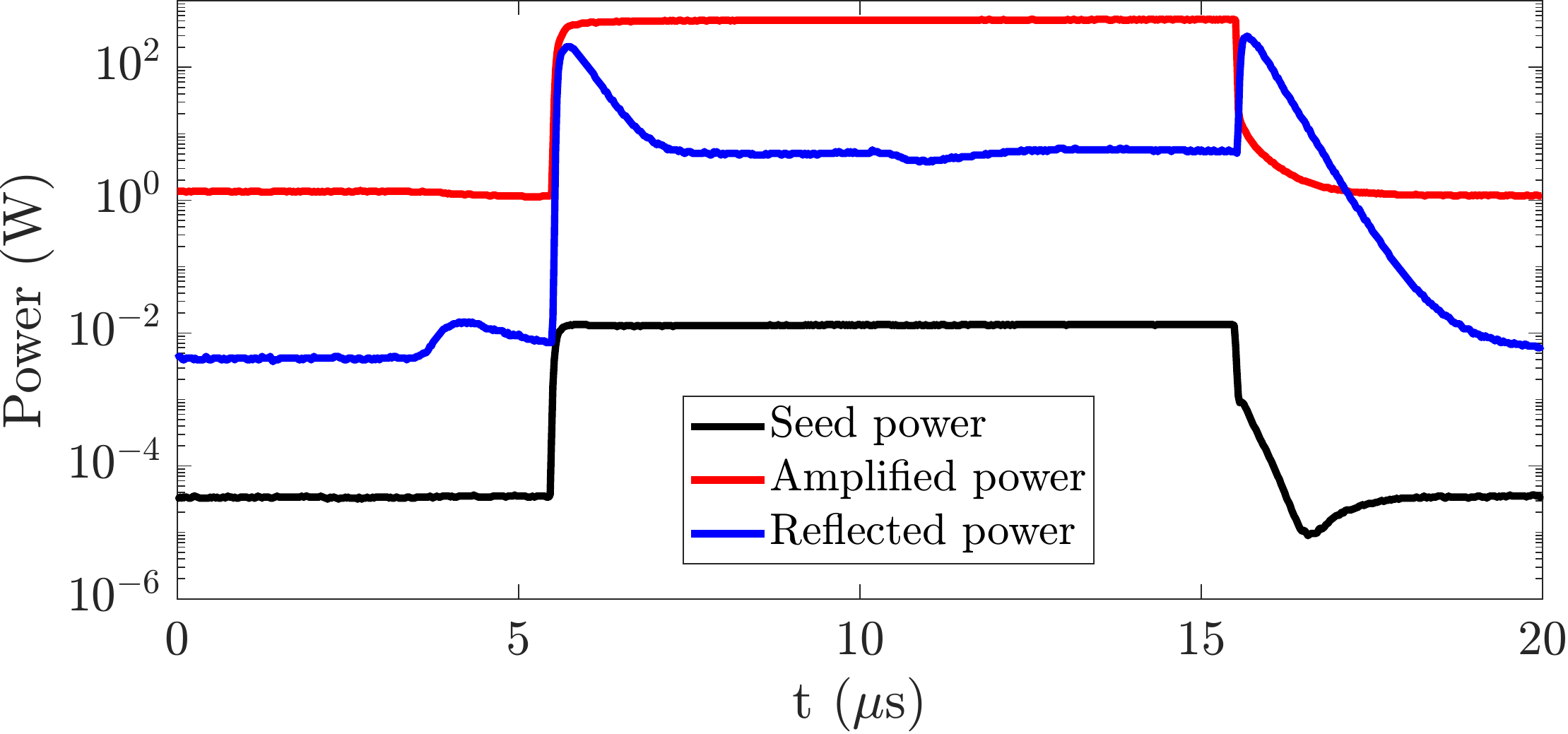}
	\caption{The signals measured by the power detectors of the seed, amplified, and reflected RF signal.}
	\label{fig:DVD_signals}
\end{figure}
\begin{figure}[t!]
	\centering
	\includegraphics[width=1\linewidth]{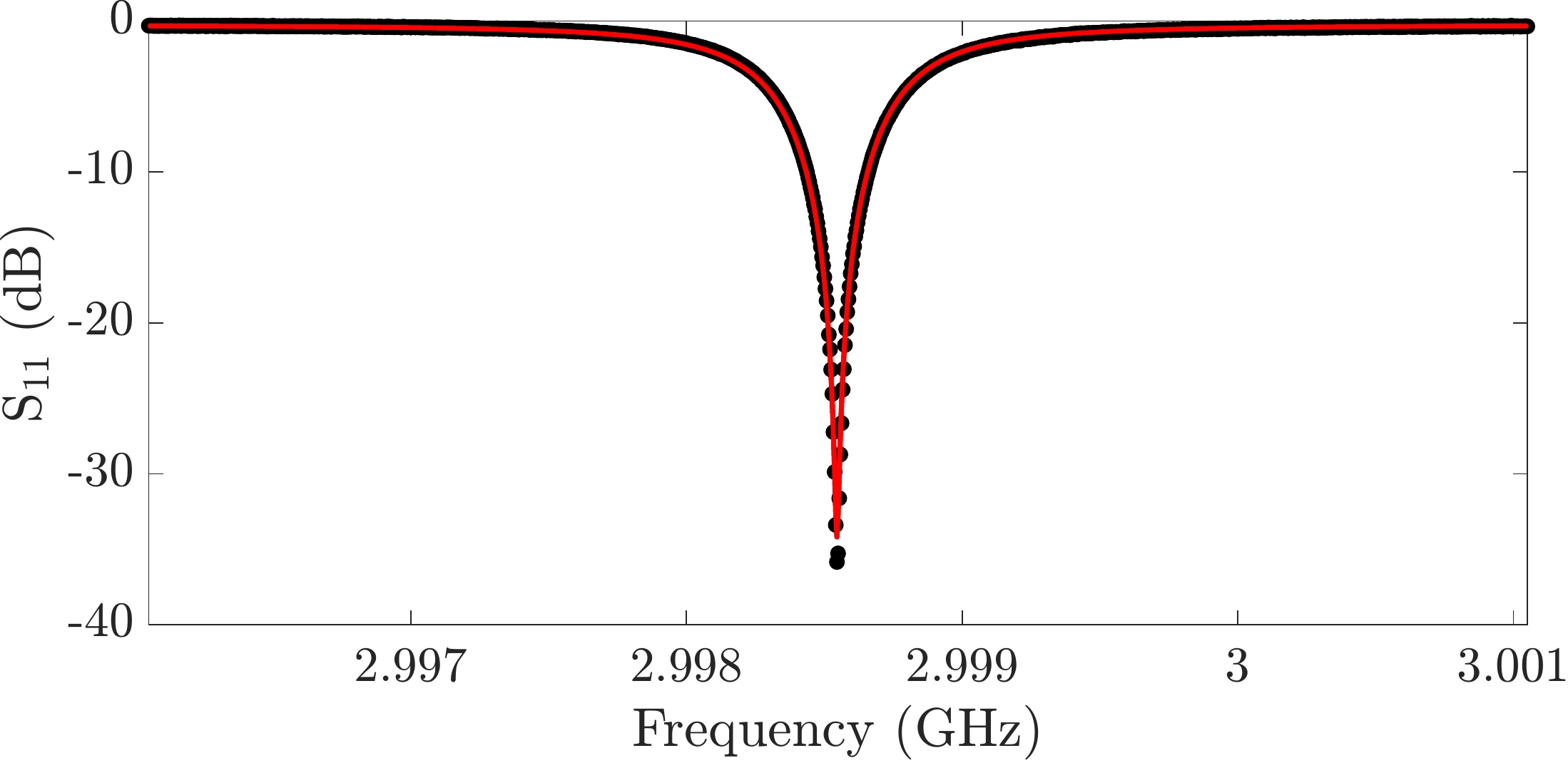}
	\caption{$S_{11}$ (reflection) plot of the RF cavity under vacuum at a temperature of 40.2 $^{\circ}$C, the black points are the obtained data and the solid red line a fit according to Eq. (\ref{eq:S11_fit}).}
	\label{fig:cavity_reflection}
\end{figure}

Through the use of a Si-PIN AEPX65 photodiode and a constant fraction discriminator system, a transistor-transistor logic (TTL) signal is generated from a split-off portion of the oscillator output. This signal is used to phase-lock a 3 GHz signal to the laser oscillator. The RF pulses are created in a custom-made 19" case, a schematic overview of its internals are shown in Fig. \ref{fig:DVD} where the elements in the transparent red box are housed in the 19" case. This signal is delivered to the 19" RF rack where the method of operation is as follows: the CW signal is first modulated in phase and amplitude using the PS6000L IQ-modulator,  after which it is amplified (ZX60-LN33+). It is then used as an input for an RF-switch (ZYSWA-2-50DR+) which is triggered with a 10 \textmu s TTL signal generated by a SpinCore Pulseblaster (SP17 model: PB12-100-4k). The output of the switch is a 10 \textmu s duration 2.99855 GHz RF signal which is henceforth called the `seed' signal. A Narda 3003-20 19.2 dB directional splitter sends the split signal to a power detector (Mini-Circuits ZX47-40-S+) after which it is sampled by an oscilloscope. The bulk of the signal is sent to a pulsed kW amplifier (microwave amps AM84-3S2-50-60R) which is triggered by the same Spincore Pulseblaster PPG with a 20 \textmu s TTL signal (5 \textmu s rise and fall time). This solid-state (GaAs FET) amplifier saturates at 60 dBm and can amplify the seed signal by 50 dB at repetition frequencies of 1 kHz for 10 \textmu s seed pulses (1\% duty cycle). The amplified signal re-enters the 19" case and is sent through a bi-directional coupler (MECA 722S-40-3.100) which splits off -40 dB of the input signal. This split-off signal gets attenuated further by a 20 dB attenuator (Mini-Circuits VAT-20+) after which a similar power detector converts the RF signal to a voltage signal to be processed by the oscilloscope. The bulk of the RF is now sent to the TM$_{010}$ cavity (through an N-type cable) where most of the power is absorbed, the reflected signal from the cavity is sent back to the bi-directional coupler and again -40 dB is split off, converted to a voltage signal, and sampled by the oscilloscope. All the connections between the components in the 19" case mentioned in this section are made with semi-rigid RG402 cable.
\\
\indent The signals generated by the power detectors are used as diagnostic tools to determine the amount of power delivered to and reflected from the accelerator structure. Additionally, when the cavity is turned off, a `transient' can be observed in the reflected power signal from which the quality factor of the cavity can be calculated during operation \citep{Gyure_2015}.
\\
\indent Fig. \ref{fig:DVD_signals} shows the converted voltage signals obtained from the RF infrastructure 19" case for the seed, amplified, and reflected power signal.
These signals were obtained at full forward power. 
\\
\indent The reflection of the structure ($S_{11}$) at a temperature of 40.2 $^{\circ}$C was measured with a network analyzer (Rohde and Schwarz ZVH8, calibrated with a ZV-Z121 50$\Omega$ kit) and is shown in Fig. \ref{fig:cavity_reflection}. The data points are converted to the corresponding voltage reflection coefficient $\left|\Gamma\right|$ and fitted according to \citep{Slater_1969, Ninhuijs_2021}:
\begin{equation}\label{eq:S11_fit}
	\left|\Gamma\right|^{2}\equiv\frac{P_{r}}{P_{0}}=
	\frac{\left(\sigma-1\right)^{2}\left(\beta\delta\right)^{2}+\left(\sigma+\beta-1\right)^{2}}
	{\left(\sigma+1\right)^{2}\left(\beta\delta\right)^{2}+\left(\sigma+\beta+1\right)^{2}},	
\end{equation}
where $P_{r}$ is the measured reflected power, $P_{0}$ the input power, $\sigma$ the reciprocal of the voltage standing wave ratio (VSWR), $\beta$ the coupling coefficient of the structure, and $\delta\approx\frac{2Q_{0}}{\beta}\left(\frac{f-f_{0}}{f_{0}}\right)$ with $f_{0}$ the resonant frequency of the structure. The data is then converted back to its $S_{11}$ form along with the fit. 

\subsection{Validation of experimental approximations}

\begin{figure}[t!]
	\centering
	\includegraphics[width=1\linewidth]{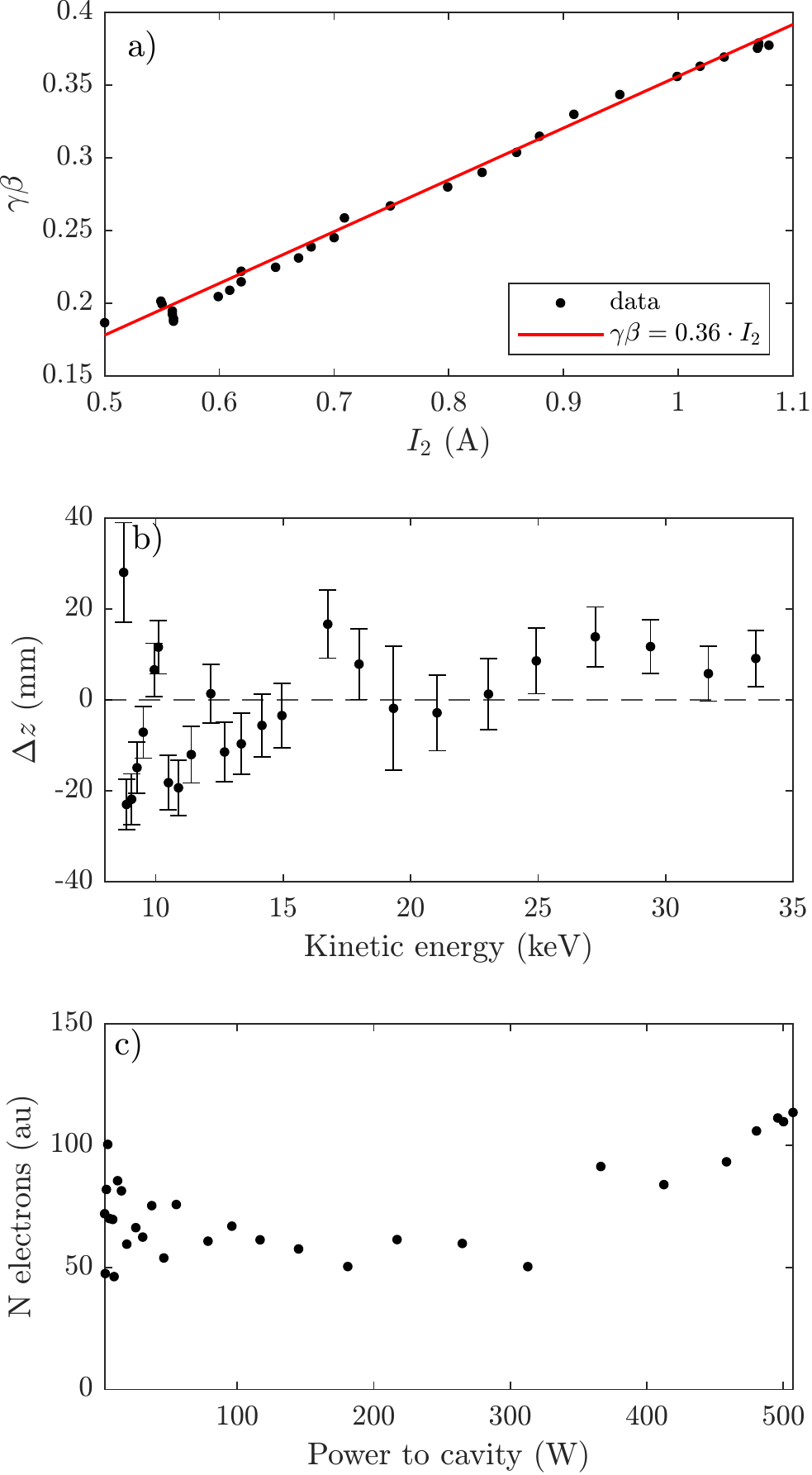}
	\caption{The longitudinal momentum, expressed in terms of $\gamma\beta$ as a function of the current supplied to the focusing solenoid with the red line a linear fit between the black data points (a). Simulation results showing the longitudinal position of the transverse waist with respect to the Au sample position (indicated by the dashed black line) as a function of the kinetic energy of the bunch, the errorbars indicate the location where the transverse rms bunch size increased to $\sqrt{2}$ times the rms waist size (b). The amount of electrons incident on the detector as a function of the power supplied to the RF structure (c).}
	\label{fig:GPT_quality_checks}
\end{figure}

\indent For each measurement the electron beam is focused on the Au sample by varying the current through the second focusing solenoid. This is done by first finding the rms spot size on the detector when the beam is not additionally accelerated and a waist-scan is performed to locate the waist on the sample holder. The Au sample was then inserted and the resulting spot size of the direct beam noted in order to re-obtain through visual inspection the proper location of the transverse waist after changing the rf amplitude. After the measurements it was found that the transverse spot sizes of the direct beam were on average $80(94)\pm17(22)$ \textmu m on the detector for the $x(y)$ direction with all RF power settings considered. This large spread in measured spot sizes implies that the transverse waist position changed as the RF power was varied to the structure, the extent of which will be considered through detailed particle tracking simulations.
\\
\indent Among the experimental variables, the power delivered to the RF structure and the current supplied to the second focusing solenoid have significant influence on the location of the transverse waist. From theory it is known that the focal length $f_{\textnormal{sol}}$ of a focusing solenoid is proportional to: $f_{\textnormal{sol}}\propto\left(\frac{\gamma\beta}{I_{\textnormal{sol}}}\right)^{2}$ with $I_{\textnormal{sol}}$ the current supplied to it and $\gamma\beta$ the longitudinal momentum of the electron bunch. Verifying the linear relationship between these quantities gives an indication on how well the transverse focal plane has been aligned with the sample plane for various kinetic energies. Fig. \ref{fig:GPT_quality_checks}(a) shows the experimentally obtained $\gamma\beta$ versus the focusing solenoid currents, a good correspondence with the expected behavior is found.
\\
\indent With the focusing solenoid currents known and the forward/reflected RF powers obtained from the RF infrastructure, detailed simulations can be done in order to check if the transverse waist of the electron bunch is located on the sample plane with these parameters. The results of these simulations, all conducted with bunches containing 500 electrons are shown in Fig. \ref{fig:GPT_quality_checks}(b). From here it is found that the longitudinal position of the transverse waist oscillates around the location of the Au sample with an rms fluctuation of $\sim13$ mm. The error bars give the distance in which the rms waist increases by a factor $\sqrt{2}$, analogous to the Rayleigh length in Gaussian optics.
\\
\indent Finally, the amount of electrons per pulse incident on the detector is plotted as a function of the power supplied to the RF cavity, shown in Fig. \ref{fig:GPT_quality_checks}(c). This electron signal is dependent upon the kinetic energy of the bunch, creating more charge carriers at the detector as the bunch energy increases, this is taken into account when calculating $N_{e}$. It is found that the amount of electrons reaching the detector stays roughly the same for all RF powers delivered to the RF structure.


\end{document}